\documentclass[
aps,%
12pt,%
final,%
notitlepage,%
oneside,%
onecolumn,%
nobibnotes,%
nofootinbib,%
superscriptaddress,%
noshowpacs,%
centertags]%
{revtex4}

\def\eps{\epsilon}

\def\gev{\,{\rm GeV}}
\usepackage{epsfig}
\begin{document}
%\selectlanguage{english}
\title{Extraction of information on transversity GPDs from $\pi^0$ and $\eta$ production on EIC of China}

\author{\firstname{Ya-Ping}~\surname{Xie}}
\email[]{xieyaping@impcas.ac.cn}
\affiliation{Institute of Modern Physics, Chinese Academy of Sciences, Lanzhou 730000, China}
\affiliation{University of Chinese Academy of Sciences, Beijing 100049, China}

\author{\firstname{S.~V.}~\surname{Goloskokov}}
\email[]{goloskkv@theor.jinr.ru}
\affiliation{
Bogoliubov Laboratory of Theoretical Physics, Joint Institute for Nuclear Research, Dubna 141980, Moscow region, Russia}

\author{\firstname{Xurong}~\surname{Chen}}
\email[]{xchen@impcas.ac.cn}
\affiliation{Institute of Modern Physics, Chinese Academy of Sciences, Lanzhou 730000, China}
\affiliation{University of Chinese Academy of Sciences, Beijing 100049, China}
\affiliation{South China Normal University, Guangzhou 510006, China}

\begin{abstract}
The General Parton Distributions (GPDs) are applied to study the hard
Pseudoscalar Meson Production (PMP)  at high energies. The PMP
amplitudes are be obtained within the GPDs factorization. They are expressed in terms of
GPDs convolution functions, which are most essential in PMP reactions. We
show that these convolution functions can be extracted from the PMP data in
future EIC of China (EicC). Predictions of $\pi^0$ and $\eta$ production at typical EicC
energies are performed.
\end{abstract}
\maketitle

\section{Introduction}\label{intro}
General Parton Distributions (GPDs) are important to study the nucleon structure\cite{ji1,Radyushkin:1996ru}.
There are a couple processes to study GPDs, for example, Deep Virtual Compton Scattering(DVCS) \cite{dvcs},
Hard Exclusive Meson Production(HEMP)\cite{dvmp1, dvmp2, gk06, gk09}, Time-like Compton Scattering (TCS)\cite{Berger:2001xd}
and Hard Exclusive Neutrino Production (HENP)\cite{Pire:2017tvv}. In HEMP
process, vector meson and pseudoscalar meson (PM) producttion can be studied using GPDs factorization. Vector
meson production is employed in twist-2 GPD \cite{gk06} and PM production is adopted in
twist-3 GPD \cite{gk09}.

Handbag approach is a valid method adopting GPDs to perform PM production in photon-nucleon
scattering. In handbag approach, the helicity amplitudes are constructed in GPDs convolution functions.
Employing the helicity amplitudes, the differential cross sections can be obtained in expression of convolution
functions of GPDs.

The COMPASS and CLAS collaborations have measured $\pi^0$ and $\eta$ differential cross
sections\cite{compass, compass02, clas}. On the other side, the cross sections of $\pi^0$ and $\eta$ can
be calculated in handbag approach applying GPDs.
Then, we can compare the theoretical predictions and the experimental data.

 In the future,
The US and China will build Electron-Ions Colliders (EICs) to investigate nucleon structure \cite{eic, eicc}. GPD is a important
aspect to study for EIC in US (EIC-US) and EIC in China(EicC). We will perform $\pi^0$ and $\eta$ prediction at EicC kinematics regions.

The convolution functions of GPDs are important in PM production. The GPD convolution
functions can be extracted from the PM differential cross sections. In this work, $E_T$ convolutions as a functions of W will be
extracted and exhibited at EicC energy regions.

This paper is organized as follow, the second section is introducing the theoretical formulas. The numerical
results and conclusion will be presented in Sec. III.
\section{Theoretical frame}
In handbag approach, the differential cross sections can be obtained using helicity amplitudes.
The helicity amplitudes can be expressed in GPD convolutions. Hence, the GPD convolutions
are important in the handbag approach.
The unpolarized differential cross sections  $e p\to eM p$ (M is PM)
can be separated into a couple of partial cross sections which are observables of
the process $\gamma^*p\to Mp$. The unpolarized differential cross section of
$\gamma p\to M p$ is given as
%\cite{Goloskokov:2022rtb,Goloskokov:2022mdn}.
\begin{eqnarray}\label{partial-cross-sections}
\frac{d^2\sigma}{dtd\phi} &=& \frac{1}{2 \pi}
(\frac{d\sigma_T}{dt} +\eps \frac{d\sigma_L}{dt}
       + \eps\cos{2\phi}\,\frac{d\sigma_{TT}}{dt}
             +\sqrt{2\eps(1+\eps)}\cos{\phi}\frac{d\sigma_{LT}}{dt})\,.
\end{eqnarray}
Here $\eps = (1-y)/(1-y+y^2/2)$.
The partial differential cross sections are obtained adopting the $\gamma^* p\to M
p$ helicity amplitudes.  The leading twist-2 $\sigma_L$ differential cross sections and
 twist-3 $\sigma_T$ are obtained via helicity amplitudes as follow
\begin{eqnarray}\label{ds}
\frac{d\sigma_L}{dt} &=& \frac{1}{\kappa}
[\mid {M}_{0+,0+}\mid^2 +\mid {M}_{0-,0+}\mid^2],\quad
\frac{d\sigma_T}{dt}=  \frac{1}{2 \kappa}(\mid {
	M}_{0-,++}\mid^2  +2 \mid {M}_{0+,++}\mid^2)\,.
\end{eqnarray}
 The $\sigma_{LT}$ and $\sigma_{TT}$ cross sections have twist-3 nature  as well
\begin{eqnarray}\label{ds2}
\frac{d\sigma_{LT}}{dt} &=& -\frac{1}{\sqrt{2} \kappa} {\rm
Re}\Big[{M^*}_{0-,++}{M}_{0-,0+}\Big],\quad
\frac{d\sigma_{TT}}{dt} = -\frac{1}{\kappa} \mid {M}_{0+,++}
\mid^2.
\end{eqnarray}

The twist-2 amplitudes can be expressed via convolution functions of twist-2 GPDs.They are given as
\begin{eqnarray}\label{conv}
{M}_{0-,0+}=\frac{e_0}{Q}\frac{\sqrt{-t'}}{2m}\langle \tilde E\rangle,\quad
{M}_{0+,0+}=\sqrt{1-\xi^2}\frac{e_0}{Q}[\langle \tilde H\rangle -
\frac{\xi^2}{1-\xi^2}\langle \tilde E\rangle].
\end{eqnarray}
The twist-3 amplitude can be obtained from twist-3 GPDs convolutions as
\begin{eqnarray}
{M}_{0-,++}= \frac{e_0}{Q}\sqrt{1-\xi^2}\langle {H_T}\rangle,\quad
{M}_{0+,++}= -\frac{e_0}{Q}\frac{\sqrt{-t'}}{4m}\langle {\bar
E_T}\rangle.
\end{eqnarray}
The detail information of the amplitudes can refer to Refs. \cite{gk09,gk11,Goloskokov:2022rtb,Goloskokov:2022mdn}.

The GPD convolutions are important in handbag approach. It is calculated as integration of hard
amplitude and GPD functions. The transversity convolutions reads as:
\begin{equation}\label{ht}
\langle H_T\rangle =\int_{-1}^1 dx
{\cal H}_{0-,++}(x,...)\,H_T;\;
\langle \bar E_T\rangle =\int_{-1}^1 dx
{\cal H}_{0-,++}(x,...)\; \bar E_T.
\end{equation}
Here ${\cal H}_{0-,++}(x,...)$ is a hard scattering part. There is a parameter $\mu_P$ in twist-3 PM wave function.
It is large and enhanced by the chiral condensate. In this work, we apply $\mu_P = 2\gev$  at scale of 2\gev.

The GPDs are constructed adopting the double distribution
representation \cite{mus99}
\begin{equation}
  F(x,\xi,t) =  \int_{-1}
     ^{1}\, d\rho \int_{-1+|\rho|}
     ^{1-|\rho|}\, d\gamma \delta(\rho+ \xi \, \gamma - x)
\, \omega(\rho,\gamma,t),
\end{equation}
which connects GPDs $F$ with Parton Distribution Functions (PDFs) $h$ with the double
distribution function $\omega$. For the valence quark double distribution, it is given as
\begin{equation}\label{ddf}
\omega(\rho,\gamma,t)= h(\rho,t)\,
                   \frac{3}{4}\,
                   \frac{[(1-|\rho|)^2-\gamma^2]}
                           {(1-|\rho|)^{3}}.
\end{equation}
 The  $t$- dependence in PDFs $h$ is expressed in the Regge form
\begin{equation}\label{pdfpar}
h(\rho,t)= N\,e^{(b-\alpha' \ln{\rho})
t}\rho^{-\alpha(0)}\,(1-\rho)^{\beta},
\end{equation}
and $\alpha(t)=\alpha(0)+\alpha' t$ is the corresponding Regge
trajectory. The parameters in Eq.~(\ref{pdfpar}) are determined from
the known information of CTEQ6 PDFs \cite{CTEQ6} e.g, or from the
nucleon form factor analysis \cite{pauli}.

 The $H_T$ GPDs are expressed with transversity PDFs as follow
\begin{equation}
  h_T(\rho,0)= \delta(\rho);\;\;\; \mbox{and}\;\;\;
\delta(\rho)=N_T\, \rho^{1/2}\, (1-\rho)\,[q(\rho)+\Delta
q(\rho)],
\end{equation}
 by adopting the model \cite{ans}. We define $t$ -dependence of $h_T$  as
 in Eq.~(\ref{pdfpar}).
There are several parameter sets of GK model. two sets parameter which are labeled
Model-II and Model-III are listed in Table.~\ref{parameters}. In the calculations of this work,
the results will be labeled as Model-II and Model-III.

\section{Numerical results and conclusion}
\begin{figure}
	\centering
	\includegraphics[width=5in]{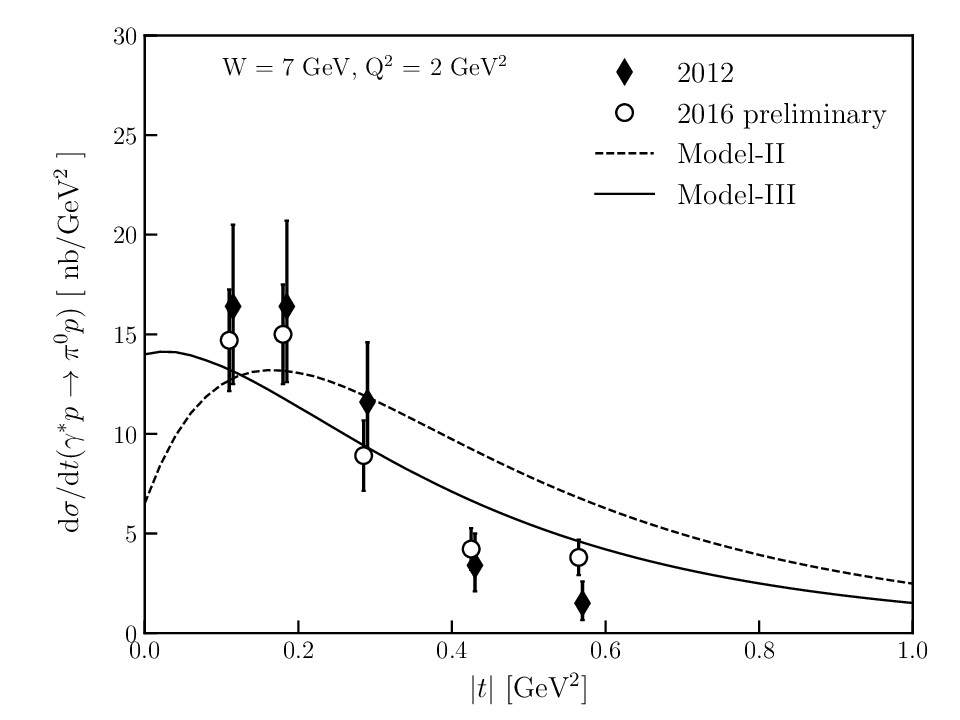}
	\caption{Models results at COMPASS kinematics.
		Experimental data are taken from 2012\cite{compass} and 2016 preliminary\cite{compass02}, dashed line represents
		the results of Model-II and solid curve indicates the prediction of Model-III.}
	\label{Fig01:diagram}
\end{figure}

\begin{figure}
	\centering
	\includegraphics[width=5in]{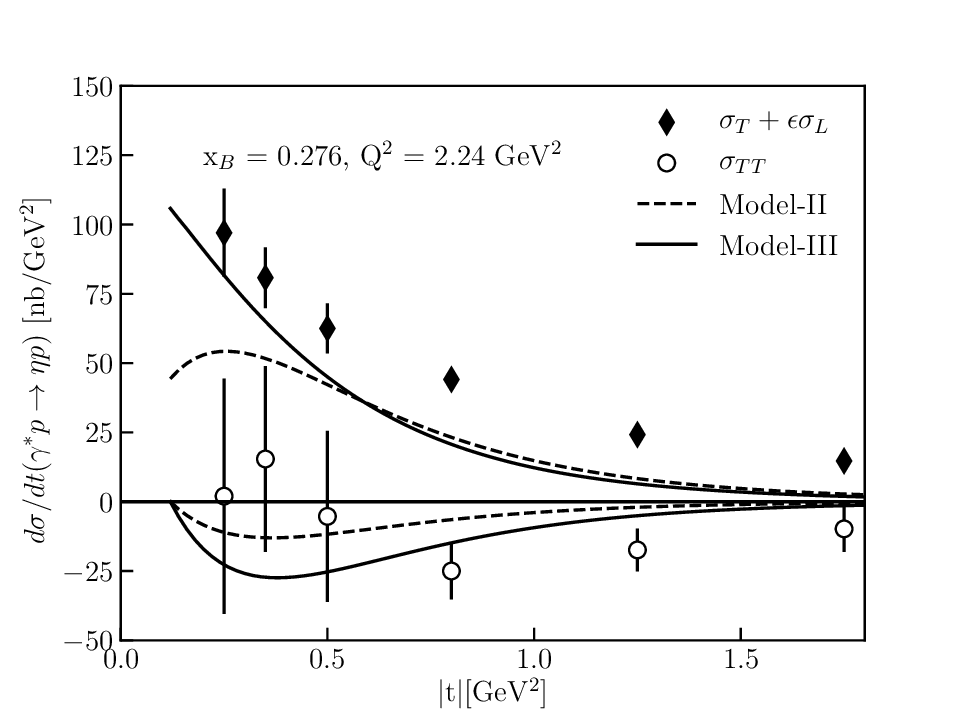}
	\caption{Cross sections of $\eta$ production in the CLAS energy
		range together with the data \cite{clas}. Upper graph (larger than zero) is for $\sigma$
		and lower graph (less than zero) is for $\sigma_{TT}$. The dashed curve is
		Model-II and the solid curve is Model-III.}
	\label{Fig02:diagram}
\end{figure}

Adopting the frame in the previous section, we can calculate the differential
cross section of PM in handbag approach within GPDs. GK model
of GPDs will be performed in this work. In our analyses \cite{Goloskokov:2022rtb,Goloskokov:2022mdn},
There were three sets
of parameterization for GK model. The first set used in \cite{Goloskokov:2022rtb} is
closed to model used in \cite{gk11}. It describes well CLAS data but at COMPASS
energies \cite{compass} gives cross sections two times larger with respect to
experiment data\cite{Goloskokov:2022rtb}. To describe CLAS and COMPASS data simultaneously, we
apply Model-II and III \cite{krollpr} used in \cite{Goloskokov:2022mdn}. Models results at
COMPASS energy are exhibited in Fig.~\ref{Fig01:diagram}. It can be seen that Model-III better describes
COMPASS data\cite{compass} at $|t'|>$0.2 GeV$^2$. Now we have preliminary COMPASS data \cite{compass02}, that are quite
close to Model-III results. The models prediction for $\eta$ production at CLAS are presented in
Fig.~\ref{Fig02:diagram}. It is found that models give some differences in cross sections at low momenta
transfer.

\begin{figure}
	\centering
	\includegraphics[width=5in]{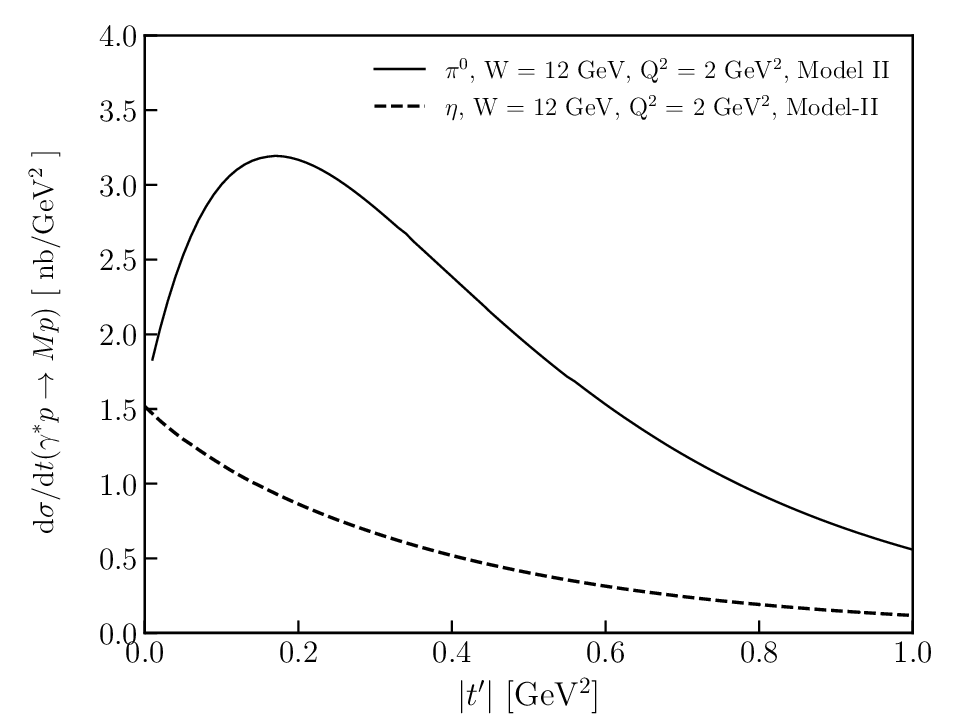}
	\caption{Differential cross sections as a function of $|t'|$ for $\pi^0$ and $\eta$ production at EicC energy.
	 The solid curve is $\pi^0$ and dashed curve is $\eta$ cross sections.}
	\label{Fig03:diagram}
\end{figure}

In Fig.~\ref{Fig03:diagram}, we presented our Model-II results for $\pi^0$ and $\eta$ production at energy
typical for China EicC collider (we presented only one model results). Cross section for $\pi^0$ production is larger with respect to
$\eta$ by factor about 3 at $|t'|>$0.2 GeV$^2$. The same property is observed at low
CLAS energies.

\begin{figure}
	\centering
	\includegraphics[width=5in]{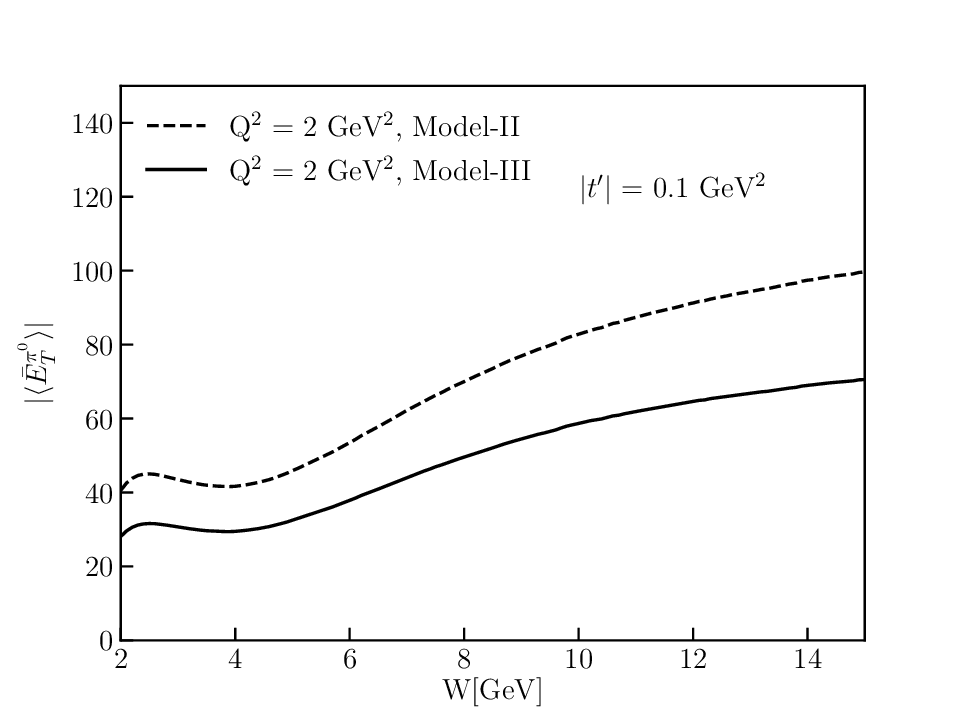}
	\caption{Extracted from the cross section transversity convolution functions as
		a function of W for $\pi^0$  production at $|t'|$= 0.1 GeV$^2$. The dashed curve is
	 Model-II and the solid curve is Model-III.}
	\label{Fig04:diagram}
\end{figure}

Using $\sigma_T$ and $\sigma_{TT}$ cross section from Eq.~(2) and Eq.~(3), one can extract
$\langle \bar E_T\rangle$ and $\langle H_T\rangle$ convolutions Eq.~(5) \cite{conv} from the
experimental data on $\pi_0$ and $\eta$ production. The experimental data on
these reactions will be available when EicC will operate. To demonstrate
such possibility instead realistic data we use here our model results for the cross
section. In Fig.~\ref{Fig04:diagram} we show $W$ dependence of $\langle \bar E_T\rangle$ at fixed
momentum transfer and $Q^2$ for $\pi_0$ and $\eta$ cases. Similar results can be
found for $\langle H_T\rangle$ which is smaller with respect to $\langle \bar
E_T\rangle$. Using flavor decompositions for $\pi^0$ and $\eta$ GPDs,one can
extract transversity convolutions for $u$ and $d$ quarks, see e.g. \cite{Goloskokov:2022mdn}.
Note
that in most experiments, the unseparated cross section $\sigma=\sigma_T+\epsilon\sigma_L$
can be measured. We show that the transversity dominance $\sigma_T \gg \sigma_L$
is valid from CLAS till EicC energies. This means that instead
$\sigma_T$, experimental results for $\sigma$ can be used. This analyses can help us to study
the nature of the GPDs convolutions.

\section*{Acknowledgment}
S.G. expresses his gratitude to P.Kroll for long-time
collaboration on GPDs study. The work is partially supported by the CAS president's international fellowship initiative (Grant No. 2021VMA0005)
and Strategic Priority Research Program of Chinese Academy of Sciences (Grant NO. XDB34030301) .

\newpage

\begin{table*}[t]
	\begin{center}
		\begin{tabular}{| p{1.8cm}| p{1.cm} | p{1.0cm} | p{1.0cm} | p{1.0cm} | p{2cm} | p{2cm} | p{1.3cm} | p{1.3cm} |}
			\hline
			Model & GPD & $\alpha(0)$ & $\beta^u$& $\beta^d$&  $\alpha^\prime
			[\gev^{-2}]$ & $b [\gev^{-2}]$ & $N^u$ &
			$N^d$ \\
			\hline
			Model-II &
			$\widetilde{E}_{\rm n.p.}$ & 0.32 &4 & 5& 0.45 & 0.6 & 18.2 & 5.2 \\
			&$\bar{E}_T$& -0.1&4 & 5 & 0.45 & 0.67 & 29.23 & 21.61 \\
			&$H_T$ & -  &- & - & 0.45& 0.04 & 0.68 & -0.186 \\
			\hline				
			Model-III	&$\bar{E}_T$& -0.1 &4 & 5&  0.45 & 0.77 & 20.91 & 15.46 \\
			&$H_T$ & - & -& -& 0.45 & 0.3 & 1.1 & -0.3 \\			
			\hline
		\end{tabular}
	\end{center}
	\caption{Regge parameters and normalizations of the GPDs at a
		scale of $2\,\gev$. }\label{parameters}
\end{table*}

\end{document}